\documentclass[aps,prd,showpacs,preprintnumbers,nofootinbib,twocolumn]{revtex4-1}
\usepackage[active]{srcltx}
\usepackage{bm}
\usepackage{amssymb}

\newcommand{\be}{\begin{equation}}
\newcommand{\e}{\end{equation}}
\newcommand{\bear}{\begin{eqnarray}}
\newcommand{\ear}{\end{eqnarray}}
\newcommand{\nline}{\nonumber \\}
\newcommand{\f}{\frac}

\newcommand{\de}{d}
\newcommand{\im}{i}

\begin{document}

\title{Reply to Comments by Batic et al.}

\author{T. Roy Choudhury}
\affiliation{Harish-Chandra Research Institute, Chhatnag Road, Jhusi, Allahabad 211 019, India}
\email[]{tirth@hri.res.in}
\author{T. Padmanabhan}
\affiliation{IUCAA, Ganeshkhind, Pune 411 007, India}
\email[]{nabhan@iucaa.ernet.in}

\date{\today}

\begin{abstract}

We explain why the analysis in our paper \cite{cp04} is relevant and correct.

\end{abstract}
\pacs{
04.70.-s, 
04.30.-w, 
04.62.+v
}
\maketitle

The Comment by \cite{batic} on our paper \cite{cp04} is essentially based on two claims: (i) The integrals used in \cite{cp04} 
to determine the QNMs do not exist for negative 
values of the argument, thereby invalidating the extraction of poles using 
Born approximation. (ii) The poles of the integral used in \cite{cp04} to determine
the quasinormal modes (QNMs) in the Schwarzschild case  
come not only from Gamma function but from Whittaker functions 
as well [eq (7) of \cite{batic}].  

The second claim is easy to dispose of as erroneous, which we will do first. It is straightforward to see
from the formula [9.211.4] of \cite{gr94} that 
\bear
\int_0^{\infty} \f{x^{\im \omega/\kappa}}{(x+1)^s} {\rm e}^{\im \omega x/\kappa}
&= &\Gamma\left(1+\f{\im \omega}{\kappa}\right)
\nline
&\times& 
\Psi\left(1+\f{\im \omega}{\kappa}, 2-s+\f{\im \omega}{\kappa},-\f{\im \omega}{\kappa}\right)
\nline
\ear
where $\Psi$ is the confluent hypergeometric (Tricomi) function
[also written as $U(a,b,z)$]. This function
$\Psi$ is regular for all finite $a$ and $b$. Hence, the only poles arise from
the Gamma function and no other functions are needed for determining the pole structure.

Let us now take up the first point (i) which is essentially related to the existence of integrals  of the form
[e.g., eq (12) of \cite{batic}]
\be
I = \int_0^{\infty} \de x~ x^{\im \omega/\kappa}
{\rm e}^{\im \omega x/\kappa}
\label{integral}
\e
where $\kappa = (4M)^{-1}$. If we introduce the parameters
$\nu = 1 + \im \omega/\kappa$ and $\mu = -\im \omega/\kappa$, the integral
can be evaluated as
\be
I = \int_0^{\infty} \de x~ x^{\nu - 1}
{\rm e}^{-\mu x} = \f{\Gamma(\nu)}{\mu^{\nu}};~~~
\mbox{Re}(\mu) > 0,~ \mbox{Re}(\nu) > 0
\label{eq:gammafn}
\e
The condition $\mbox{Re}(\mu) > 0$ translates to $\omega_I > 0$, which
does not affect our results as we are interested only in (large)
positive values of $\omega_I$.
If we let $z = \mu x$, then we essentially have to evaluate the Gamma function integral
\be
\mu^{\nu} I = I' = \int_0^{\infty} \de z~z^{\nu-1} {\rm e}^{-z} = \Gamma(\nu);~~~
\label{gamma}
\e
The authors claim that this evaluation of the integral as Gamma function is valid only for $Re \nu>0$ (which translates to  translates 
to $\omega_I < \kappa$) while we are interested in ref.\cite{cp04} for large $\omega_I$. 
This objection, too, is easy to take care of.

The point to note is that, $\Gamma(\nu)$ can be defined by analytic continuation for negative values of $\nu$. Even though for $\mbox{Re}(\nu) < 0$, the integrand in Eq.\ref{gamma} behaves as $\sim z^{\nu-1}$ as 
$z \to 0$  the analytic continuation allows one to define the Gamma function everywhere in complex plane. Once this is done, one can easily extract the poles. A simple, text book way to extract this result is to use the identity:
\be
\Gamma(\nu)~\Gamma(-\nu) = -\f{\pi}{\nu \sin \nu \pi}
\label{id}
\e
which provides a way to analytically continue $\Gamma(\nu)$ to negative values
of $\mbox{Re}(\nu)$. This also shows that the function has simple poles for 
 negative integral values of 
$\mbox{Re}(\nu)$ arising from the $\sin(\pi\nu)$ factor, which is also a well-known result. \textit{These
are precisely the poles that we are interested in which gives us the 
desired QNM structure.} Of course, by the very definition of a `pole', the integral diverges at the pole; we stress that the whole exercise is to determine precisely where this occurs! The integral exists in a open neighborhood of the first order pole which is what we used in our analysis.
The analytic continuation is based on the standard assumption in scattering theory
that the scattering amplitude (and hence the integral in Eq. \ref{integral}) is analytic
everywhere in the $\omega$ plane, except for a finite number of poles. The key idea developed in \cite{padmanabhan03} and \cite{mmv03} and used in the paper under discussion \cite{cp04} was to use this assumption, identify the poles and relate it to the QNM.
 
If one does not want to use the identity in Eq.\ref{id} but want to work directly with integral in Eq.\ref{gamma} and extract the information about the poles, that is also possible. We only have to treat the integral in  Eq.\ref{gamma} as a limit of a sequence of integrals with a suitable regularization parameter and study the poles. This can be done in many ways and we outline one procedure:
Consider the integral:
\begin{equation}
I' =  \int_{0}^{\infty} \de z~z^{\nu-1} {\rm e}^{-z}~{\rm e}^{-a/z}
= 2 a^{\nu/2} K_{-\nu}(2 \sqrt{a})
\end{equation}
which is well defined even for $\mbox{Re}(\nu) < 0$ because of the regulator factor $e^{(-a/z)}$. (Here $K_{\nu}$ is the modified Bessel function and  the relation 
 can be 
obtained from \cite{gr94}, Sec 8.40-8.43.) We treat the integral in Eq.\ref{gamma} (especially for $\mbox{Re}(\nu) < 0$)
as the limit of $I'$ when $a$ is a positive infinitesimal quantity. This gives (again using Sec 8.40-8.43 of \cite{gr94} and interpreting $a \to 0^+$ as a positive infinitesimal value for the regulator):
\bear
I' &=& 
\lim_{a \to 0^+} \int_{0}^{\infty} \de z~z^{\nu-1} {\rm e}^{-z}~{\rm e}^{-a/z}
\nline
&=& \lim_{a \to 0^+} 2 a^{\nu/2} K_{-\nu}(2 \sqrt{a})
\nline
&=&  \Gamma(\nu) + \lim_{a \to 0^+} a^{\nu} \left[\Gamma(-\nu) + {\cal O}(a)
\right] 
\ear
Note that the last equality in the above equation is valid only when $\nu$ is 
{\it not} a integer. In fact, if we take the limit $\nu \to -n$ (where $n$ is positive integer), 
the integral diverges \textit{as it should}, because our previous analysis using Eq.\ref{id} has already told us that the integral has simple poles at $\nu= -n$. We can determine the 
nature of the singularity arising from these poles trivially. When $\nu \to -n$, 
we obtain
\bear
\lim_{\nu \to -n} (\nu + n) I' &=& \lim_{\nu \to -n} (\nu + n) \Gamma(\nu)
\nline
&+& \lim_{a \to 0^+} \lim_{\nu \to -n} a^{\nu} \left[(\nu + n) \Gamma(-\nu) + {\cal O}(a) \right]
\nline
&=& \f{(-1)^n}{n!} + \lim_{a \to 0^+} a^{-n} \left[...\right] \neq 0
\ear
which shows that the singularity of $I'$ at $\nu = -n$ is a simple pole for finite regulator. 
(The procedure is very similar to the $i\epsilon$ prescription used in field theoretic calculations.) 
Same arguments are valid for
the integral (18) of \cite{batic}. We stress that the integrals are not expected to exist for 
$\nu = -n$, which are precisely the poles we want to determine! What we need is a sensible definition of the integral in the open neighborhood of the poles --- which can be provided in many ways, of which we have described two.

Finally, we would like to point out that paper in question which is being commented upon
\cite{cp04} is a follow-up of two earlier papers \cite{padmanabhan03} and \cite{mmv03} developing the same technique and containing the same integrals. 
It is somewhat surprising that the authors of \cite{batic} decided to comment 
 a third, follow-up paper rather than the first two! We did not discuss the details
 of the regularization, analytic continuation etc. in our work \cite{cp04} as the basic ideas were already implicit in the previous
 published work.

\end{document}